\documentclass[doublecol]{epl2}
\usepackage{amsmath}
\usepackage{graphicx}
\usepackage{color}
\usepackage{hyperref}
\usepackage{psfrag}
\usepackage{stmaryrd}
\usepackage{amssymb}
\usepackage{wasysym}
\usepackage[latin1]{inputenc}

\def \bs{\boldsymbol}

\def \be{\begin{equation}}
\def \ee{\end{equation}}
\def \ba{\begin{eqnarray}}
\def \ea{\end{eqnarray}}
\def \wt{\widetilde}
\def \wh{\widehat}
\def \mc{\mathcal}
\def \bse{\begin{subequations}}
\def \ese{\end{subequations}}

\title{Counter-rotation in an orbitally shaken glass of beer}
\shorttitle{Counter-rotation in an orbitally shaken glass of beer}

\author{F. Moisy\inst{1}\thanks{E-mail: \email{moisy@fast.u-psud.fr}}
\and J. Bouvard\inst{1}
\and W. Herreman\inst{2}}
\shortauthor{F. Moisy \etal}

\institute{\inst{1} Laboratoire FAST, Universit\'e Paris-Sud, CNRS, Universit\'e Paris-Saclay, France \\
\inst{2} Laboratoire LIMSI, Universit\'e Paris-Sud, CNRS, Universit\'e Paris-Saclay, France}

\pacs{47.32.-y}{Vortex dynamics; rotating fluids}
\pacs{47.35.-i}{Hydrodynamic waves}

\abstract{Swirling a glass of wine induces a rotating gravity wave along with a mean flow rotating in the direction of the applied swirl. Surprisingly, when the liquid is covered by a floating cohesive material, for instance a thin layer of foam in a glass of beer, the mean rotation at the surface can reverse. This intriguing counter-rotation can also be observed with coffee cream, tea scum, cohesive powder, provided that the wave amplitude is small and the surface covering fraction is large. Here we show that the mechanism for counter-rotation is a fluid analog of the rolling without slipping motion of a planetary gear train: for sufficiently large density, the covered surface behaves as a rigid raft transported by the rotating sloshing wave, and friction with the near-wall low-velocity fluid produces a negative torque which can overcome the positive Stokes drift rotation induced by the wave. }

\begin{document}

\maketitle

\section{Introduction} \label{sec:intro}

The mean flow induced by a rotating sloshing wave in an orbitally shaken cylinder partially filled with liquid consists in a global rotation in the direction of the applied swirl, along with toroidal recirculation vortices~\cite{Prandtl1949,Hutton1964,Reclari2014,Bouvard2017,Timokha2017}. This mean flow, commonly observed when swirling a glass of wine, is essential for mixing processes such as in bioreactors for the cultivation of biological cells~\cite{Kim2009,Weheliye2013}. Here we describe an intriguing and, to our knowledge, unreported phenomenon: when gently swirling a liquid covered by a floating raft of cohesive material, the mean rotation at the surface can reverse. This intriguing phenomenon is easily observed in a cup of espresso coffee or a glass of beer covered by a thin layer of foam.  It can also be observed in a cup of tea, because of the thin scum film composed of calcium and organic matter that forms at the water surface~\cite{Mossion2008}.

Nontrivial surface flows in orbital shaking strikingly illustrates the critical influence of surface contamination in wave-induced flow generation~\cite{Craik1982,Martin2006,Higuera2014,Francois2015,Perinet2017}. We show here that the reversal in the floating raft rotation results from a complex interplay between transport by the rotating sloshing wave, friction with the container wall, and internal stress in the viscoelastic raft~\cite{Langevin2014}. For a deformable raft of small extent, the Stokes drift induced by the sloshing wave dominates and the raft is in co-rotation. On the other hand, when the raft is sufficiently large and rigid, the negative frictional torque induced by the low-velocity region near the wall may overcome the Stokes drift contribution, producing counter-rotation of the raft. This mechanism can be seen as a fluid analog of the rolling without slipping motion of a planetary gear train, also observed in orbitally shaken granular media~\cite{Feltrup2009}.

\section{Experiments} \label{sec:expe}

\begin{figure}
    \centerline{\includegraphics[width=8.4cm]{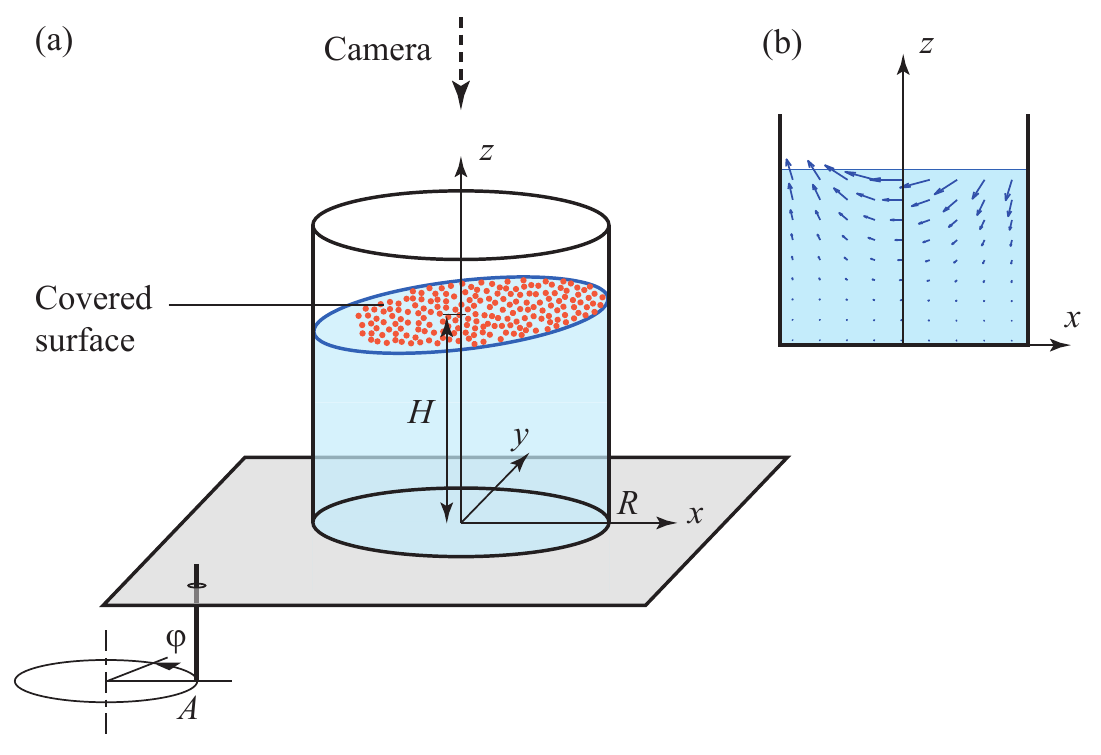}}
    \caption{(Colour Online) (a) Experimental setup.  The cylinder is orbitally shaken at a constant frequency $\Omega=d \varphi / dt$ along a circular trajectory of radius $A$, maintaining a fixed orientation with respect to an inertial frame of reference. The liquid surface is covered with foam, powder or beads (see text), and its mean rotation is visualized by a camera synchronized with the forcing frequency. (b) Wave flow in the plane $(x,z)$, from the free-surface linear potential theory,  shown at the phase $\varphi=\pi/2$.}
	\label{fig:setup}
\end{figure}

Experiments with various liquids, surface covering and cylinder size have been carried out. The experimental set-up, sketched in fig.~\ref{fig:setup}(a), is similar to the one described in Bouvard {\it et al.}~\cite{Bouvard2017}. A cylinder of radius $R$ filled up to height $H$ is orbitally shaken by an eccentric motor along a circular trajectory given by  ${\bf r}_c(t) = A (\cos \Omega t \, {\bf e}_x + \sin \Omega t \, {\bf e}_y)$. In the frame attached to the cylinder, this motion induces a rotating centrifugal force per unit mass of magnitude $A \Omega^2$, which excites a rotating gravity wave of angular phase velocity prescribed by the forcing frequency $\Omega$.  We measure the mean motion of the surface covering averaged over the wave period with a camera located above the cylinder. In order to filter out the large-amplitude wave motion and measure only the second order mean flow, the image acquisition is synchronized with the forcing~\cite{Perinet2017,Bouvard2017}. This stroboscopic measurement is sensitive to the total (Lagrangian) mass transport, which includes the (Eulerian) steady streaming contribution and the Stokes drift contribution.

\begin{figure}
\begin{center}
\includegraphics[width=7.5cm]{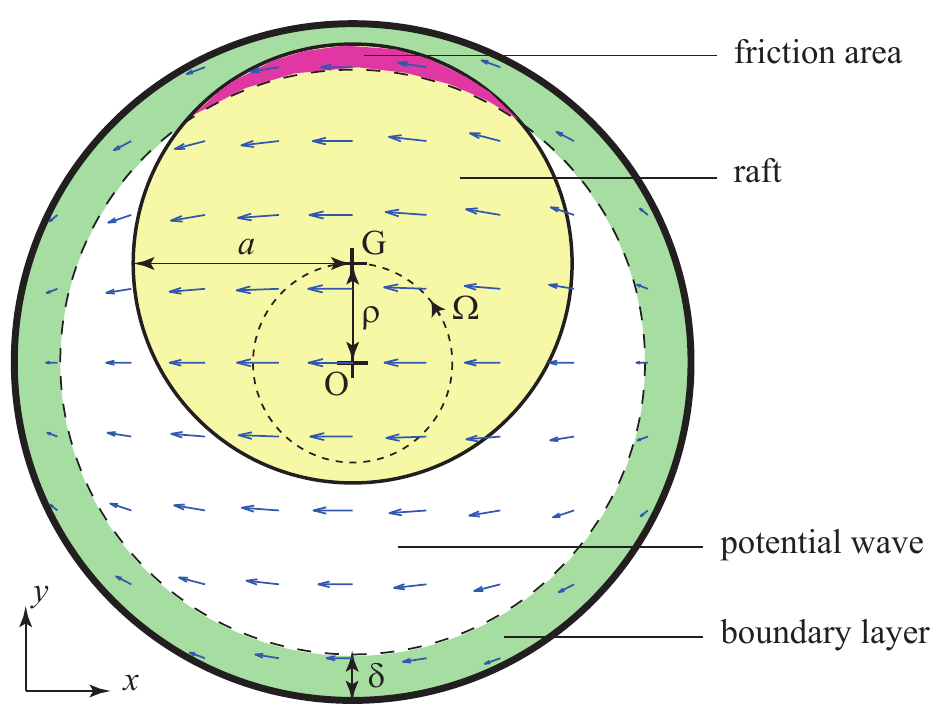}
\caption{(Colour Online) Motion of a circular raft (in yellow) floating at the surface of the liquid in the reference frame of the cylinder.  The wave velocity at the surface, depicted with blue arrows, describes closed orbits, approximately circular near the center of the cylinder, in the positive direction at the forcing frequency $\Omega$ (the wave field is shown here at the phase $\varphi = \pi/2$). The raft is transported by the wave, such that its center of mass describes a circular orbit of radius $\rho$ at angular velocity $\Omega>0$. The upper edge of the raft (in red) lies over the boundary layer near the wall (in green), inducing a negative torque and hence a counter-rotation $\omega<0$ of the raft.}
\label{fig:sk}
\end{center}
\end{figure}

We first briefly recall the classical orbital sloshing flow in the case of a free surface. According to the linear potential theory (recalled in section A of the Supplementary Material (SM)), for small forcing amplitude $\epsilon = A/R \ll 1$, the rotating gravity wave can be described as the superposition of two linear sloshing waves at right angle with $\pi/2$ phase shift~\cite{Ibrahim2005,Faltinsen2014,Reclari2014}.  The velocity field is sketched in fig.~\ref{fig:setup}(b) and fig.~\ref{fig:sk} at a particular phase of the forcing ($\varphi = \pi/2$), such that the cylinder velocity $d {\bf r}_c / dt$ is along $-{\bf e}_x$; each vector arrow describes a circle, nearly horizontal at the center and nearly vertical near the cylinder wall. The key non-dimensional number in this problem is
\begin{equation}
\chi = \frac{\epsilon}{(\omega_{1}/\Omega)^2 - 1},
\label{eq:scaluw}
\end{equation}
where
$$
\omega_1^2 = \frac{g k_1}{R} \tanh(k_1 H/R)
$$
is the fundamental resonance frequency of the cylinder, and $k_1 \simeq 1.841$ is the first zero of the derivative of $J_1$, the Bessel's function of first kind and first order.   In the validity range of the potential theory ($\chi \ll 1$), the wave flow is linear in $\chi$: the wave velocity is  $u \simeq \chi \Omega R$, and the surface elevation of the fluid, which also sets the radius of the particle orbits near the surface, is $\rho \simeq \chi R$. Nonlinear interactions of this rotating gravity wave induce a mean flow. In the weakly nonlinear regime, this mean flow is expected to be quadratic in wave amplitude~\cite{Longuet1953,Batchelor1967,Riley2001}: $\overline{u} \simeq \chi^2 \Omega R$. The two scaling laws, $u \simeq \chi \Omega R$ and $\overline{u} \simeq \chi^2 \Omega R$, have been recently confirmed in experiments with free surface over the range $\chi \simeq 10^{-2} - 10^{-1}$, in which both $\epsilon$ and $\Omega/\omega_1$ were varied~\cite{Bouvard2017}.

Mean flows generated by propagating waves in containers include in general both an Eulerian steady streaming contribution, driven by the oscillating boundary layers, and a Lagrangian Stokes drift contribution\cite{Stokes1847,Longuet1953,Batchelor1967,Riley2001,Nicolas2003}. In the orbital sloshing problem, the mean flow consists in a robust central rotation at angular velocity $\overline{\omega}_0 / \Omega \simeq \chi^2$, with weak dependence on the fluid viscosity, and poloidal recirculations of weaker amplitude, mostly active near the contact line~\cite{Bouvard2017}. Analysis suggests that the central rotation is dominated by the Stokes drift induced by the quasi-inviscid rotating wave, while the poloidal recirculations are dominated by steady streaming. Importantly, in the range of wave amplitude $\chi \simeq 10^{-2} - 10^{-1}$ explored here and in Ref.~\cite{Bouvard2017}, in the case of a free surface, the mean central rotation is always in the direction of the wave ($\overline{\omega}_0 / \Omega > 0$). Any counter-rotating motion of the surface must therefore result from a modification of the mean flow by the surface covering.

\begin{figure}
    \centerline{\includegraphics[width=7.3cm]{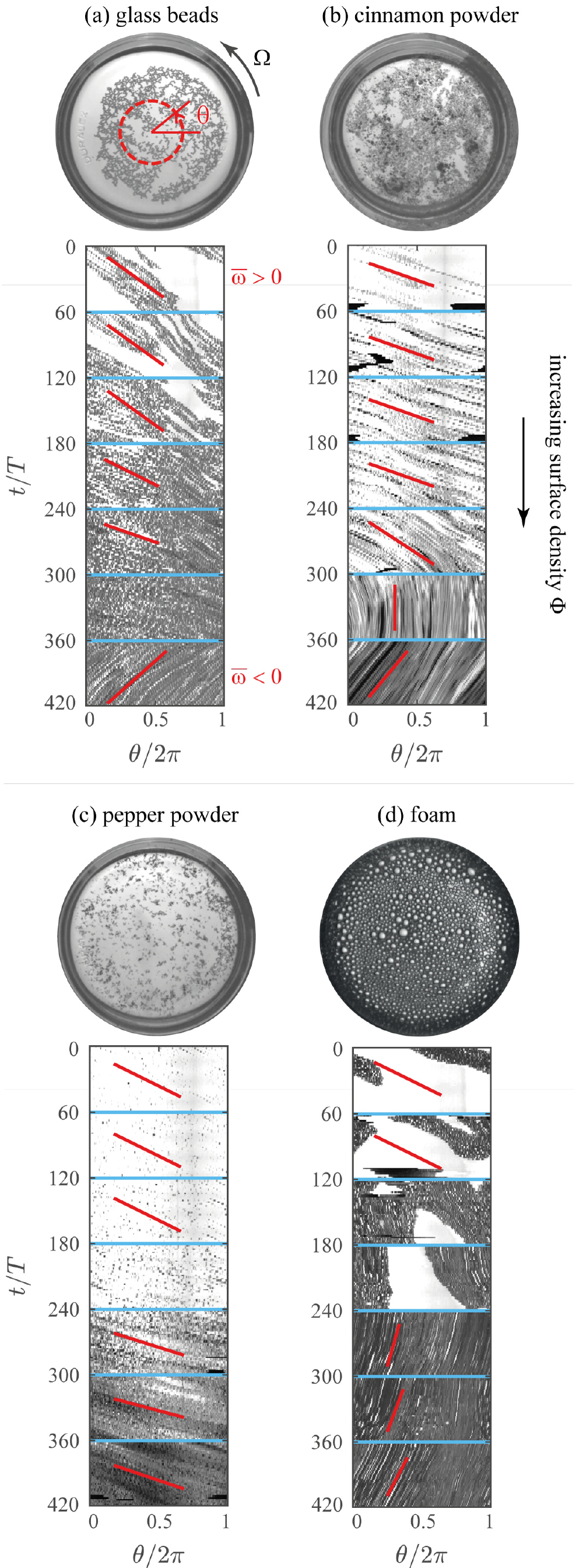}}
    \caption{(Colour Online) Spatio-temporal diagrams of the surface covering along the angular coordinate $\theta$ sampled along a circle of radius $r=14$~mm (red dashed circle), showing the direction of rotation as the surface density is increased (see Supplementary Movies supplied in ref. \cite{SM}). The blue lines indicate times at which the density is increased, by pouring additional material on the surface. The red segments show the angular velocity $\omega$ of the pattern, evolving from co-rotation at small time (small density) to counter-rotation at large time (except in the case c, which remains in co-rotation for all density).}
		\label{fig:spatio}
\end{figure}

We illustrate now the effect of the surface covering on the direction of the mean flow. A series of experiments using water with various surface coverage is shown in fig.~\ref{fig:spatio}: (a) glass beads, 0.5 mm in diameter; (b) cinnamon powder; (c) pepper powder; (d) foam.  For each surface coverage, the forcing frequency is kept constant, while the surface density is gradually increased by simply pouring additional material with a spoon (see Supplementary Movies supplied in ref. \cite{SM}). These experiments are performed in a cylindrical container, 37.5~mm in radius, filled up to height $H=20$~mm (resonance frequency $\omega_1 = 182$~rpm). The forcing amplitude is $\epsilon=A/R = 0.035$ and the forcing frequency $\Omega/\omega_1 = 0.77$. The normalized wave amplitude, $\chi = 0.048$, lies in the weakly nonlinear range $\chi \simeq 10^{-2} - 10^{-1}$ for which a co-rotating mean flow $\overline{\omega}_0 / \Omega \simeq \chi^2 >0$ is observed in the absence of surface coverage.

The rotation of the surface pattern for the four types of covering is visualized in fig.~\ref{fig:spatio} using spatio-temporal diagrams: at each forcing period the pattern is sampled along the angular coordinate $\theta$ of a centered circle of radius $r=14$~mm (see the red dashed circle in fig.~\ref{fig:spatio}(a)). In all cases, the pattern makes a complete rotation in typically 100 forcing periods, {\it i.e.} $|\overline \omega| / \Omega \simeq 0.01$. All the coverings show co-rotation at low surface density, but only the cases (a), (b), (d) (glass beads, cinnamon powder and foam) turn to counter-rotation at large density, while the pepper (c) remains in co-rotation at all density. The key difference between the coverings is that they all form a cohesive raft at the surface of the liquid except the pepper powder (c). In the cases of glass beads (a) and foam (d), cohesion of the raft is due to the attractive capillary forces, an effect sometimes referred to as ``Cheerios effect''~\cite{Vella2005}. In the case of the cinnamon powder (b), cohesion is due to the release of a surfactant layer showing strong surface elasticity. On the other hand, the surfactant layer released by the pepper powder (c) turns out to induce a strong repulsive force between the grains, which prevents the cohesion of the raft.   Note that the glass beads and the cinnamon powder rafts remain approximately circular and centered, whereas the raft of bubbles tends to migrate and spread along the wall because of the strong attraction of the meniscus.

These first experiments indicate that a necessary condition for counter-rotation is the formation of a coherent raft of sufficient size and rigidity. Such raft behaves as a two-dimensional elastic solid, able to transmit shear stresses applied at its periphery through force chains~\cite{Vella2004}. This suggests the following picture for the transition to counter-rotation. At low surface density, the raft is small and is simply transported by the rotating gravity wave: its center is in translation along a circular orbit of radius $\rho \simeq \chi R$ at frequency $\Omega$, with a slow second-order solid-body rotation $\overline \omega>0$. Far from the boundaries, this second-order rotation is dominated by the Stokes drift contribution~\cite{Bouvard2017}. The steady-streaming contribution, mostly active near the contact line, mainly corresponds to poloidal recirculation vortices: it moves the raft away or towards the center of the cylinder, without changing significantly its angular velocity. As the surface density is increased, the raft becomes larger, so that the region of its edge that is closer to the wall, where the wave is the highest, experiences friction with the slower fluid (see fig.~\ref{fig:sk}).  This slower fluid region may correspond to the Stokes boundary layer, of typical thickness $\delta = \sqrt{\nu / \Omega}$, in the cases (a) and (b), or may be due to the presence of bubbles trapped in the meniscus near the wall in the case (d). Because of the raft rigidity, the resulting negative frictional torque is transmitted to the entire raft (except for the pepper powder), yielding a negative angular velocity: the raft ``rolls'' along the cylinder wall, like a planetary gear train, except that the counter-rotation rate here, $|\overline \omega|/\Omega \simeq 10^{-2}$, is much smaller than that of a solid rolling without sliding, $|\omega| / \Omega = R/a \simeq O(1)$ (with $a$ the disk radius).

\section{Regime diagram} \label{sec:reg}

\begin{figure}
	\centerline{\includegraphics[width=8.8cm]{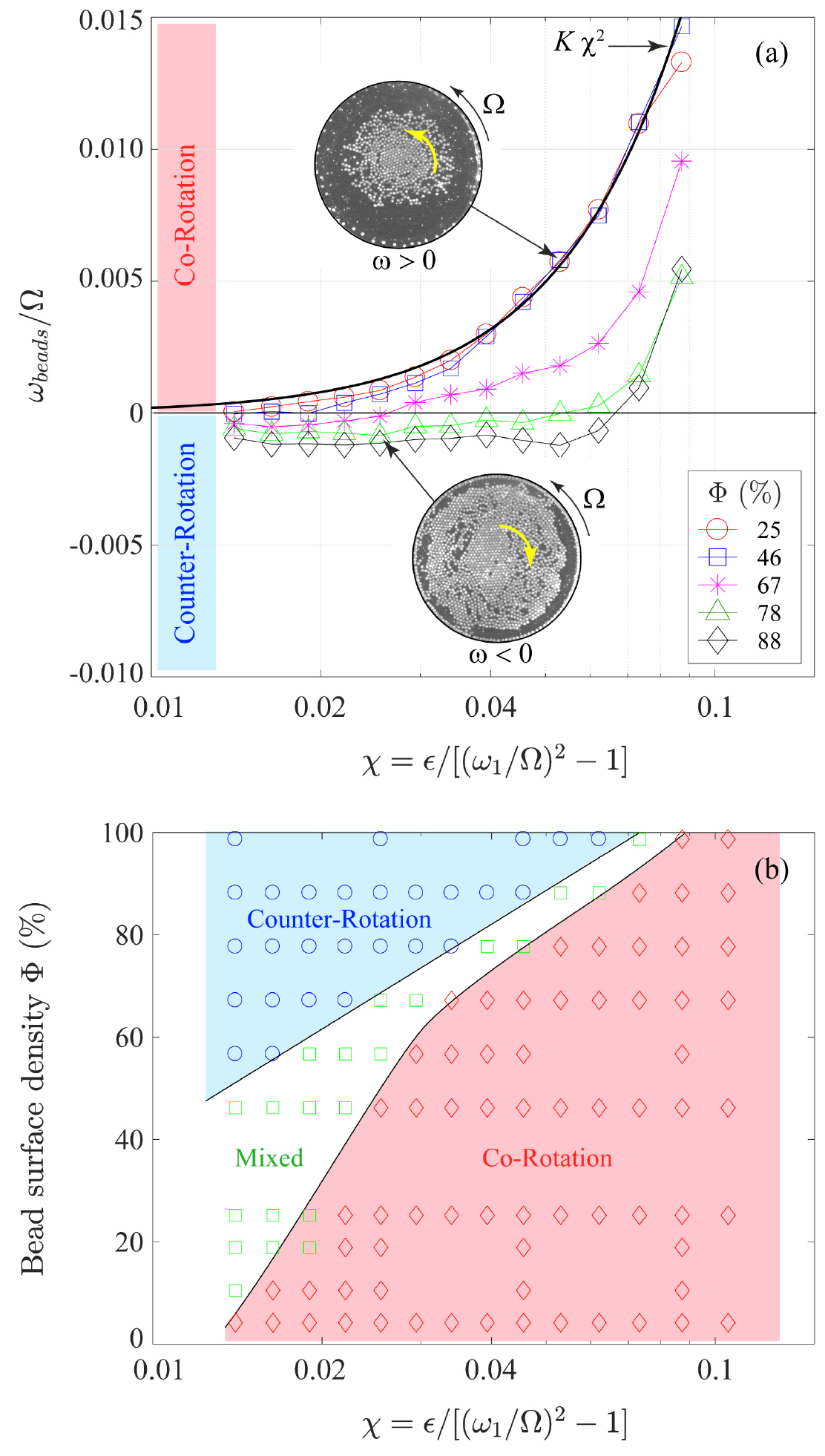}}
    \caption{(Colour Online) (a) Normalized angular velocity of the raft of beads near the center as a function of the normalized wave amplitude $\chi$ for different surface coverage density $\Phi$. At small density the beads rotate according to the Stokes drift law $K \chi^2$, with $K \simeq 2$ (black line).  (b) Regime diagram showing the sign of the angular velocity in the plane ($\chi,\Phi$).}
		\label{fig:diag}
\end{figure}

In order to describe quantitatively the transition from co- to counter-rotation, we have performed a series of experiments with a well controlled surface covering. We use $N$ polypropylene beads, of density $\rho_s = 0.90$~g/cm$^3$ and diameter $b=2$~mm, floating at the surface of silicon oil, of density $\rho = 0.95$~g/cm$^3$ and kinematic viscosity $\nu = 50$~mm$^2$~s$^{-1}$. Because of the weak density contrast and of the good wetting of oil on polypropylene, the beads float just below the surface, with an almost flat meniscus inducing a weak attractive capillary force, resulting in a relatively fragile raft. The raft formed by the beads is approximately circular and centered, with some beads trapped in the meniscus (see the insets in fig.~\ref{fig:diag}a). For the sake of comparison, the parameters in this experiment are the same as in Ref.~\cite{Bouvard2017}: the cylinder, of radius $R=51.2$~mm, is filled up to height $H=111$~mm (resonance frequency $\omega_1=180$~rpm), and shaken with forcing amplitude $\epsilon = A/R = 0.057$. The forcing frequency $\Omega$ is varied between 80 and 145~rpm, corresponding to a normalized wave amplitude $\chi$ in the range 0.014 - 0.11.  We define the bead surface density as
$$
\Phi = N \frac{\sqrt{3}}{2\pi} \left(\frac{b}{R}\right)^2,
$$
normalized such that $\Phi=100\%$ corresponds to the maximum circle packing density, obtained for $N=2380$ beads. 

Figure~\ref{fig:diag}(a) shows the normalized angular velocity of the raft of beads, $\omega_{beads} / \Omega$, as a function of $\chi$ for different bead density $\Phi$. The angular velocity $\omega_{beads}$ is determined using stroboscopic particle image velocimetry (i.e., from correlation of images separated by one forcing period), and is defined as half the mean vorticity in a centered disk of radius $R/3$, averaged over 200 forcing periods. At small $\Phi$, the bead raft is always in corotation, and follows essentially the Stokes drift induced by the rotating gravity wave: its angular velocity is well described by the law
\begin{equation}
\frac{\omega_{Sto}}{\Omega} = K \chi^2
\label{eq:ksto}
\end{equation}
(black line), with $K \simeq 2.0 \pm 0.2$. As the bead density $\Phi$ is increased, the angular velocity decreases and eventually becomes negative for moderate wave amplitude $\chi$.  In this counter-rotating regime, the normalized angular velocity $\omega_{beads} / \Omega$ shows a weak dependence with $\chi$, followed by a sharp transition to the Stokes drift co-rotation regime for $\chi > 0.06$.

The sign of $\omega_{beads}$ as a function of $(\chi, \Phi)$ is summarized in fig.~\ref{fig:diag}(b). The co-rotating region (large $\chi$ and small $\Phi$, in red) and the counter-rotating region (small $\chi$ and large $\Phi$, in blue) are sepated by a mixed regime (in green), showing both co-rotation near the center and counter-rotation near the periphery.   As $\chi$ is increased, in spite of the larger friction area with the near-wall low-velocity region due to the larger gyration radius $\rho \simeq \chi R$ (see fig.~\ref{fig:sk}), the bead raft tends to rotate in the positive direction, suggesting that the effect of the Stokes drift increases with $\chi$ more rapidly than the near-wall friction. For $\chi > 0.1$ (limit of validity of the weakly nonlinear regime), we find a positive rotation of the raft for all beads density $\Phi$.  This systematic co-rotation may be explained by a loss of coherence of the raft when transported by a too strong wave flow. The raft remains coherent if the differential drag force induced by the wave on the beads, of order $\Delta F \simeq \eta b \Delta u$ (with $b$ the bead diameter and $\eta$ the fluid viscosity) remains smaller than the capillary force $F_c$ between the beads. Taking $\Delta u \simeq b | \nabla u | \simeq b \chi \Omega$ for the velocity difference between two beads separated by a distance of order $b$, this suggests that the cohesion of the raft is lost for $\Omega \chi > F_c / (\eta b^2)$. Beyond this limit, the beads are essentially independent and locally follow the co-rotating Stokes drift induced by the wave.

\section{Model} \label{sec:model}

We propose here a model for the transition from co- to counter- rotation as the raft size is increased for small wave amplitude $\chi$, assuming that the raft remains cohesive. We model the raft as a set of $N$ floating, attractive and inertialess particles. The position vector of a particle $i$ is decomposed as $\mathbf{r}_i = \wh{\mathbf{r}}_i + z_i \bs{e}_z$, where we use hats for the horizontal components.  We suppose that the raft easily deforms in the vertical direction, so that the presence of the raft does not significantly alter the wave flow. The vertical position of the floating particles is then fixed by $z_i \simeq \eta (\wh{\mathbf{r}}_i  , t)$ where $\eta$ denotes the surface elevation associated to the wave.  We also suppose that the raft is sufficiently stiff in the horizontal direction, which means that the horizontal motion of the particles in the raft is essentially solid. A particle $i$ in the raft has horizontal velocity
\be
\frac{d \wh{\mathbf{r}}_i }{ d t } \simeq \frac{d \, \wh{\mathbf{r}}_{g}}{d t} + \omega\, \mathbf{e}_z     \times (\wh{\mathbf{r}}_{i}  - \wh{\mathbf{r}}_{g}   )  . 
\ee 
The horizontal motion of the raft is entirely characterized by its center of mass $\wh{\mathbf{r}}_g (t) = \sum_i \wh{\mathbf{r}}_i  / N$  and  by its rotation rate $\omega (t) $. 

In section B of the SM, we derive equations for $\wh{\mathbf{r}}_{g}$ and $ \omega$ starting from the fundamental force balance on the $N$ particles. In the continuum limit, we model the raft as a circular disk of radius $a$, and find
\bse \label{eq:transetrot3}
\ba
\frac{d \, \wh{\bf r}_g}{dt}  &=&  \frac{1}{\pi a^2}   \iint_{\mc{D}(t) } \wh{\bf u}(\wh{\bf r} + \bs{\eta} ,t)  \, d^2 \wh{\mathbf{r}}  \label{eq:trans3}  \\
 \omega  & =&  \frac{2}{\pi a^4}     \iint_{\mc{D}(t)}  [ (\wh{\mathbf{r}} - \wh{\mathbf{r}}_g )   \times  \wh{\bf u}(\wh{\bf r}+ \bs{\eta} ,t)  ] \cdot \mathbf{e}_z  \, d^2 \wh{\mathbf{r}} \label{eq:rot3} ,
\ea
\ese
where $\mc{D}(t)$ is the domain $ || \wh{\mathbf{r}}  - \wh{\mathbf{r}}_g || \leq a$,  $\bs{\eta}= \eta \bs{e}_z$ and $\wh{\bf u}$ is the horizontal component of the fluid flow. These equations can be used to calculate the gyration of the raft (oscillatory motion of $\wh{\bf r}_g$) and the transition from co-rotation to counter-rotation. This is done analytically in sections C, D, E of the SM using perturbative expansions in orders of $\chi$.

The first order calculation uses the potential wave flow $\wh{\mathbf{u}} \simeq \wh{\nabla} \phi$ and  simplifies  $ \wh{\bf u}(\wh{\bf r} + \bs{\eta} ,t)  \simeq  \wh{\bf u}(\wh{\bf r} ,t)$ and $\wh{\mathbf{r}} - \wh{\mathbf{r}}_g  \simeq \wh{\mathbf{r}} $  in \eqref{eq:transetrot3}. For the translational motion, we find $\wh{\bf r}_g = \rho (\cos \Omega t {\bf e}_x + \sin \Omega t {\bf e}_y)$,  with gyration radius 
\be
\rho = \chi R \underbrace{\frac{2}{k_1^2-1} \frac{R}{a} \frac{J_1(k_1 a/R)}{J_1(k_1)}}_{C(a/R)}.
\ee
The function $C$ slightly decreases with $a/R$, with $C \simeq 1.32$ for $a/R \ll 1$ and $C\simeq 0.84$ for $a/R=1$. Interestingly, for such potential wave, we find $\omega =0$: the raft cannot rotate at first order in $\chi$. This can be seen in  \eqref{eq:rot3} when using the identity $\wh{\bf r} \times \wh{\nabla} \phi = - \wh{\nabla} \times (\wh{\bf r} \, \phi)$ and the Stokes theorem. A raft rotation necessarily results from a higher  order effect, or from the presence of vorticity in the carrying wave flow.

To find the slow co-rotation at next order in wave amplitude $\chi$, we need to consider that the wave flow is modified by a steady streaming part ($ \mathbf{u} = \nabla \phi + \overline{\bf u}  $) and also, that the flow in the integrals \eqref{eq:transetrot3} is to be expressed at the moving interface. Using Taylor expansions, we express the integranda in the vicinity of $z=0$. This allows us to derive the second order formula for the time-averaged part $\overline{\omega}$ of the rotation speed: 
\be
\overline{\omega} =
  \frac{2}{\pi a^4}   \int_0^a \int_0^{2 \pi}   \left [  \wh{\overline{\bf u}}   + \overline{ ( \bs{\eta}  \cdot \nabla )  \wh{\nabla} \phi  }   \right ]_\theta    {r}^2 \, d {r} \, d {\theta}    .
\ee
Inside the brackets, we see two contributions, one due the steady streaming and one that is a Stokes drift correction.  The study of Bouvard et al. [{\rm 3}] suggested that the steady streaming flow has a weak azimuthal component ($\overline{u}_\theta \simeq 0$). Keeping only the Stokes drift contribution, we can calculate explicitly 
\be
\frac{\overline{\omega}}{\Omega}  \simeq  \chi^2  K(H/R,a/R).  \label{eq:wsto}
\ee 
The coefficient $K > 0$, given in the SM, varies from $2.97$ to $1.67$ for  $a/R$ varying from $0$ to $1$ in our set-up. We note that this interval includes the value of $K \simeq 2.0$ found experimentally (fig. 4(a)).

Finally, to find the slow counter-rotation, we must take into account that large rafts can penetrate the annular boundary layer near the cylinder wall. The fluid is slowed down there and exerts a negative torque on the rim of the raft  that can result in a counter-rotation. To model this boundary effect, we modify the potential flow as $\mathbf{u} = \nabla \phi + \mathbf{u}_{BL}$, introducing a boundary correction
\be
{\bf u}_{BL} ({\bf r},t) = - \nabla \phi ({\bf r},t) |_{r=R} \,  e^{-(R-r)/\delta},
\label{eq:ublc}
\ee
with $\delta$ a boundary layer thickness, ensuring that $\mathbf{u}$ satisfies a no-slip boundary condition. This is a very crude model of the true boundary layer structure near the moving contact line that remains intractable. We think that this simple correction is sufficient to capture the essential physics of the counter-rotation, and consider the thickness $\delta$ as an adjustable parameter.

We finally determine $\overline{\omega}_{BL}$, the counter-rotation induced by the boundary layer correction, by injecting \eqref{eq:ublc} in \eqref{eq:rot3}.
For small wave amplitude and in the limit $\rho \ll \delta$, a negative angular velocity is found as (see section E of the SM)
\be
\label{eq:wbl}
\frac{\overline{\omega_{BL}}}{\Omega} = - \chi^2 \frac{4}{k_1^2 -1} \left(\frac{R}{a} \right)^4 C(a/R) \, e^{-(R-a)/\delta}.
\ee
Interestingly, this boundary-layer contribution is of order $\chi^2$ too, because the frictional torque originates from a $O(\chi)$ velocity defect acting on a $O(\chi)$ raft displacement.  Adding the Stokes drift co-rotation (\ref{eq:wsto}) and the boundary-layer counter-rotation (\ref{eq:wbl}) finally yields the total angular velocity
\be
\label{eq:wtot}
\frac{ \overline{\omega}_{tot}}{\Omega} = \chi^2 \left[ K(a/R) -  \frac{4}{k_1^2 -1} \left(\frac{R}{a} \right)^4 C(a/R) \, e^{-(R-a)/\delta} \right].
\ee
Since $K(a/R)$ and $C(a/R)$ are slowly varying functions of order unity, we can see that the term in brackets actually changes sign for sufficiently large raft, when $R-a$ is of order of $\delta$, in agreement with the qualitative scenario proposed in the previous section.

\begin{figure}
	\centerline{\includegraphics[width=8cm]{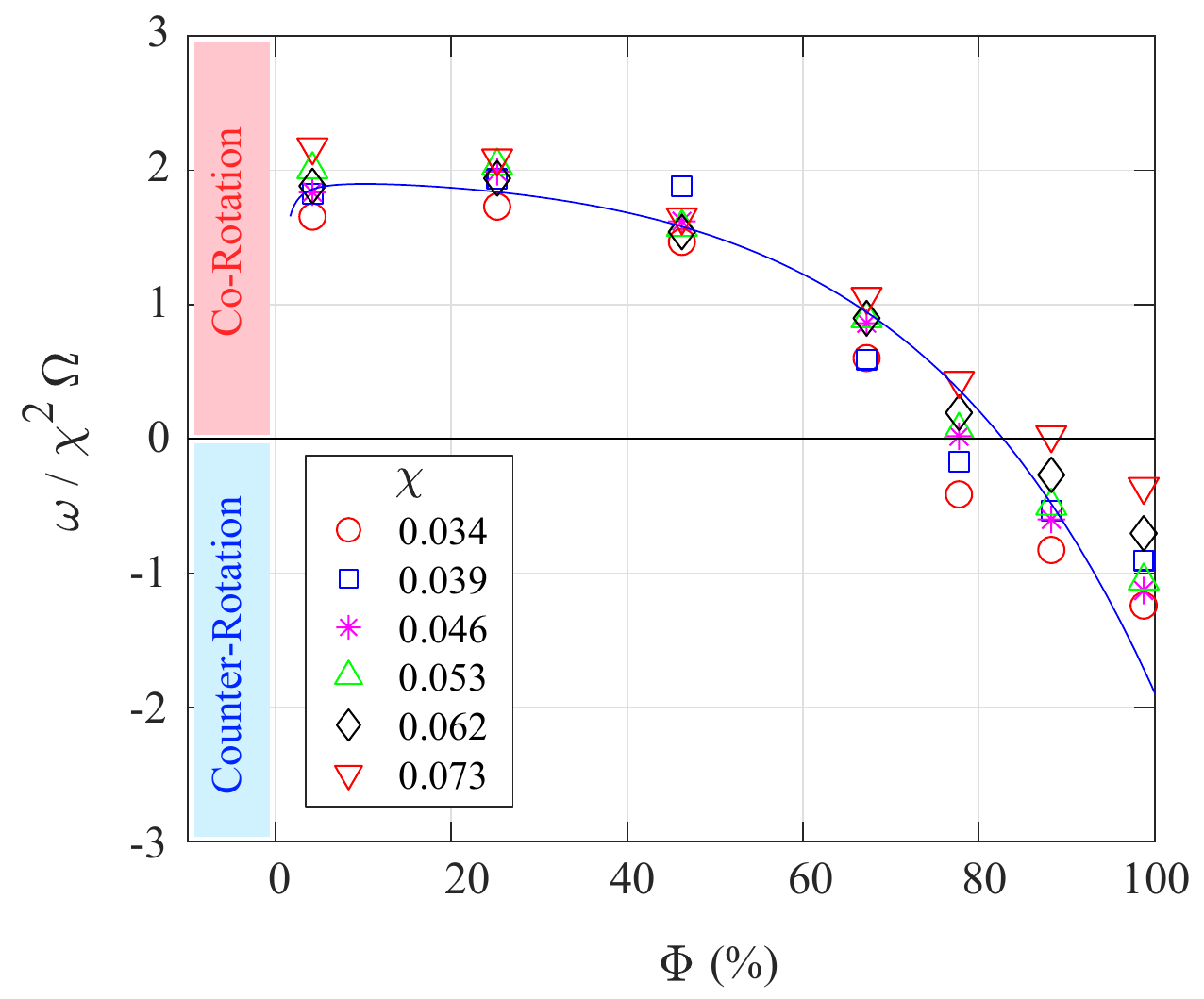}}
    \caption{(Colour Online) Normalized angular velocity of the raft as a function of the bead density $\Phi$, for wave amplitudes $\chi = 0.034-0.073$. The solid line shows the model (\ref{eq:wtot}), using a boundary layer thickness $\delta / R = 0.09$ and a surface density $\Phi = 0.74 (a/R)^2$.}
		\label{fig:wth}
\end{figure}

To provide comparison between the model (\ref{eq:wtot}) and the measured angular velocity of the bead raft, we introduce a raft compacity factor $c \leq 1$, such that the bead surface density is $\Phi = c(a/R)^2$ ($c=1$ corresponds to a raft of densily packed beads).  Figure~\ref{fig:wth} shows that a correct agreement is obtained, for a range of wave amplitude $\chi \simeq 0.034 - 0.073$. The model here is plotted for a compacity factor $c \simeq 0.74$ and a boundary layer thickness  $\delta/R \simeq 0.09$, a value of the order of the Stokes boundary layer thickness for this range of forcing frequency. For larger wave amplitude, cohesion of the raft is lost, and the measured angular velocity is larger than predicted. This confirms that the rotation rate of the raft, at least in the case of a cohesive raft, can be modeled as a balance between the positive Stokes drift induced by the rotating wave and the negative frictional torque induced by the boundary layer.

\section{Conclusion} \label{sec:concl}

The counter-rotation of a cohesive raft floating at the surface of a liquid in orbital shaking motion is a subtle phenomenon resulting from the complex interplay between wave transport, friction, and internal stress in the raft. In this paper, we show that the transition from co- to counter- rotation can be captured by a simple model, assuming a light and slightly deformable raft that does not alter the dynamics of the rotating gravity wave.  Since the model assumes a cohesive raft, it can describe only the transition from co- to counter-rotation as the raft size is increased  at moderate wave amplitude $\chi$. On the other hand, the transition from counter- to co-rototation at larger $\chi$, which relies on the loss of cohesion of the raft strained by a wave of large amplitude, cannot be captured by the present model. Note that, although the $O(\chi)$ wave flow remains essentially unaffected by the presence of the raft, the $O(\chi^2)$ mean flow in the bulk, which is driven by the mean velocity of the raft at the surface, is expected to show sign reversal too.

Using a surface covering with macroscopic material (foam, powder, beads) makes the transition to counter-rotation easy to observe with a classical laboratory orbital shaker, or even by carefully swirling the liquid by hand. As the surface covering becomes thicker, however, the feedback of the raft on the wave motion cannot be neglected. This limitation is illustrated by swirling a glass of beer with more than a few layers of bubbles: the strong dissipation induced by the foam~\cite{Sauret2015} usually prevents the onset of counter-rotation, resulting in an over-damped rotating gravity wave with no noticeable mean rotation. Inversely, the counter-rotation effect may be present even for surface contamination at the microscopic scale, hardly visible to the naked eye, for example by nanolayers of soluble organic matter such as proteins or lipids, which could lead to unexpected results when working with a supposedly free surface.

\acknowledgments

We acknowledge A. Aubertin, L. Auffray, R. Kostenko and R. Pidoux
for their experimental help, and A. Sauret for fruitful discussions.
FM acknowledges the Institut Universitaire de France.

\newpage

\onecolumn

\noindent {\Large \bf Counter-rotation in an orbitally shaken glass of beer: \\Supplementary Material}  \\

We present a theoretical model that describes the motion of a floating circular raft in the orbital sloshing problem. In section {\bf A}, we specify the orbital sloshing flow with free surface. In {\bf B}, we derive equations for the motion of the raft. In sections {\bf C} and {\bf D}, we calculate first and second order approximations of the motion of the raft to describe its gyration and its counter-rotation. In section {\bf E}, we explain the counter-rotation of the raft as a result of the interaction with the boundary layer. \\

\section{A. Flows in  an orbitally shaken cylinder}

A cylinder of radius $R$ filled with fluid up to height $H$ is being orbitally displaced as  $
{\bf r}_c = A (\cos \Omega t \, {\bf e}_x + \cos \Omega t \, {\bf e}_y ) 
$. 
Here $A$ is the amplitude of the displacement and $\Omega$ its frequency. This orbital translation drives a flow ${\bf u} (\bf{r},t)$ that we describe using cylindrical coordinates $(r,\theta,z)$ in the moving frame of reference attached to the cylinder. 

Potential flow theory provides a linear and inviscid approximation of the fluid flow [{\rm 3,4}]. For forcing frequencies $\Omega$ that are lower than the natural frequencies of the gravity waves, we have
\be \label{eq:phidef}
\bs{u} (\mathbf{r}, t ) = \nabla \phi \quad \mbox{with} \quad \phi  =   \Omega R^2  \,  \frac{2 \chi }{k_1^2 -1} \frac{J_1 (k_1 r /R) }{J_1 (k_1)} \frac{\cosh (k_1 (z+ H)/R)}{\cosh (k_1 H /R) }  \sin (\theta - \Omega t) . 
\ee 
Here $k_1 \simeq 1.841$ and 
\be
\chi  = \frac{\epsilon}{(\omega_1/ \Omega)^2 - 1}\, ,
\ee 
with $\epsilon = A/R$ and $\omega_1 = \sqrt{ g k_1 \tanh (k_1 H/R) }$ the gravity wave eigenfrequency. The surface reaches a height $z= \eta$ with  
\be
\eta =R \,  \frac{2 \chi  k_1}{k_1^2 -1} \frac{J_1 (k_1 r /R) }{J_1 (k_1)} \tanh (k_1 H/R)   \cos (\theta - \Omega t) . 
\ee
This solution only includes the dominant wave. The full solution is given in  [{\rm 3,4}].

Near the boundaries of the cylinder the inviscid potential flow model needs to be corrected in order to satisfy the no-slip boundary condition. We introduce an exponential boundary layer correction to the flow so that
\begin{equation}
{\bf u}({\bf r},t) =  {\bf \nabla \phi}  - {\bf \nabla \phi} |_{r=R} \,  e^{-(R-r)/\delta} \ .
\label{eq:mask}
\end{equation}
The boundary layer has thickness $\delta \ll R$ and $\delta$ will be a tunable parameter. This boundary layer is a very crude approximation of the real boundary layer near the contact line, but it is adequate to capture the essential physics that explains the counter-rotation.

Nonlinearities in the bulk and in the boundary layers create a weak $O(\chi^2)$ correction to the flow under the wave. A second order, more precise model of the flow in the bulk is  
\be \label{eq:flow_second}
{\bf u} (\mathbf{r}, t) = {\bf \nabla \phi} (\mathbf{r}, t) + \overline{ \bf u} (\mathbf{r} ) +  {\bf u}' (\mathbf{r} ,t ) . 
\ee
Next to the oscillatory potential wave, we find the steady streaming flow $\overline{ \bf u} (\mathbf{r} )$ as the Eulerian mean flow and some time dependent harmonics  ${\bf u}' (\mathbf{r} ,t )$. The steady streaming  flow was measured in Ref.~[{\rm 4}], but  no analytical expression is available.

\section{B. Equations of motion for the raft} 

We consider the motion of a set of $N$ identical particles of mass $m$ submerged in a fluid moving at speed $\mathbf{u} (\mathbf{r},t ) $.  The position ${\bf r}_i (t)$ and speed ${\bf v}_i (t)  = d {\bf r}_i / d t $  of a particle $i$ satisfy a fundamental force balance
\be \label{balfund}
\underbrace{m \left ( \frac{d^2  {\bf r}_i  }{d t^2 } +   \frac{d^2  {\bf r}_c    }{d t^2 } \right )}_{inertia}   =  \underbrace{\alpha_i  \left ({\bf u }({\bf r}_i,t) - \mathbf{v}_i  \right )}_{drag \ force}  + \underbrace{  \sum_{j\neq i} {\bf T}_{j \rightarrow  i}  }_{attraction}  + \underbrace{ \mc{B}_i \, \mathbf{e}_z }_{buoyancy} . 
\ee
We model the the fluid-particle interaction with a simple drag force with drag coefficients $\alpha_i$. Neighboring particles $j \neq i$ act on particle $i$ by forces ${\bf T}_{j \rightarrow i}$ that we suppose attractive and aligned with $\mathbf{r}_i - \mathbf{r}_j$. Due to gravity, there is a buoyancy term $ \mc{B}_i  \, \mathbf{e}_z$. The inertial term will be neglected in all what follows. 

We suppose that buoyancy is dominant so that all particles will remain in the immediate vicinity of the surface. Different particles are similarly submerged in the fluid, so drag coefficients should be the same for all particles: we denote $\alpha_i = \alpha$. If the particles follow the motion of the interface, we can write
\be
{\bf r}_i  =  \wh{\bf r}_i   + \bs{\eta}_i  \quad, \quad {\bf v}_i  =  \wh{\bf v}_i   + \frac{ d \bs{\eta}_i}{d t}   
\ee
with $\bs{\eta}_i  =  \eta (\wh{\bf r}_i , t ) \mathbf{e}_z  $ the surface elevation and $ \wh{\bf r}_i ,  \wh{\bf v}_i $ the horizontal components of the particle's position and speed (we use hats for horizontal field components). By writing this, we ignore dynamic feedback of the particles on the wave. The horizontal motion is constrained by  
\ba \label{balperp}
\mathbf{0}
  \simeq  \alpha \left (\wh{\bf u}({\bf r}_i,t) -   \wh{\mathbf{v}}_i  \right )  +  \sum_{j\neq i} \wh{\bf T}_{j \rightarrow i} . 
\ea
 In this balance, we suppose that the interaction forces $\sum_{j\neq i} \wh{\bf T}_{j \rightarrow i}$ are dominant so that distances $  || \wh{\mathbf{r}}_i - \wh{\mathbf{r}}_j ||  $ remain nearly fixed just as in a weakly deformable two-dimensional solid. We can decompose the particle speed as
 \be \label{approxsol}
  \wh{\mathbf{v}}_{i} \simeq \underbrace{\frac{d \, \wh{\mathbf{r}}_{g}}{d t} + \omega\, \mathbf{e}_z     \times (\wh{\mathbf{r}}_{i}  - \wh{\mathbf{r}}_{g}   ) }_{dominant \ solid \ motion} +  \underbrace{\frac{d \, \wh{\mathbf{r}}_{i}'}{d t}  }_{weak \ elastic \ motion}
\ee
separating the solid motion from a weak elastic motion. We introduce here $d \, \wh{\mathbf{r}}_{g}/ d t $, the horizontal speed of the center of mass  
$\mathbf{r}_g (t) = \sum_i \mathbf{r}_i  / N$
 and $\omega (t) $, the rotation speed of the raft. Elastic motions $d \,  \wh{\mathbf{r}}_{i}' / dt $ in the horizontal direction remain small whenever the raft is weakly compressible in the horizontal direction. To better know what this means, we estimate the order of magnitude of the elastic motion. With a fluid flow of order $\chi \Omega R$, the drag force can reach a magnitude $ \alpha \chi \Omega R $. The drag force is balanced by an elastic force that brings particles back to equilibrium positions for which $\sum_{j\neq i} \wh{\bf T}_{j \rightarrow i} = \bs{0}$. We can estimate the elastic force as  $\kappa ||   \wh{\mathbf{r}}_{i}'  ||$ with  $\kappa = || \wh{\nabla} \wh{\bf T}_{j \rightarrow i}  ||$ measuring the horizontal stiffness of the raft. The force balance leads to  $ ||   \wh{\mathbf{r}}_{i}'   ||  = \alpha \chi \Omega R  / \kappa$ as an order of magnitude for the elastic deviations and  to $ ||  d \, \wh{\mathbf{r}}_{i}' / d t  || \sim \alpha \chi \Omega^2 R  / \kappa $ for the elastic motion. Elastic motion can be ignored with respect to solid motion of order $||d \wh{\mathbf{r}}_{g}  / d t ||  \sim \chi \Omega R $ whenever 
\be
\frac{\alpha \Omega }{ \kappa}  \ll 1 . 
\ee
The stiffer the raft in the horizontal direction (the higher $\kappa$ ), the smaller the elastic motion. We suppose that this condition is fulfilled and this allows us to ignore the weak elastic motion in all what follows. 

Without elastic deviations, the motion of the raft is entirely determined by $\wh{\mathbf{r}}_g (t)$ and $\omega (t)$ for which we can derive two simple equations. 
Summing \eqref{balperp}  and  $(\wh{\mathbf{r}}_i - \wh{\mathbf{r}}_g) \times $ \eqref{balperp}  over all $N$ particles, we can identify that
\bse
\ba
\frac{d \, \wh{\bf r}_g}{dt}  &=& \frac{1}{N}  \sum_i \wh{\bf u} (\wh{\bf r}_i + \bs{\eta}_i,t)   \label{eq:trans}  \\
\omega  &=& \frac{ \sum_i [ (\wh{\mathbf{r}}_i - \wh{\mathbf{r}}_g)   \times \wh{\bf u}( \wh{\bf r}_i + \bs{\eta}_i ,t) ] \cdot \mathbf{e}_z}{ \sum_i || \wh{\mathbf{r}}_i - \wh{\mathbf{r}}_g   ||^2   }. \label{eq:rot}  
\ea
\ese
Due to Newton's third law ($\mathbf{T}_{j\rightarrow i} +\mathbf{T}_{i \rightarrow j} = \bs{0}$) and collinearity  of $\mathbf{T}_{j\rightarrow i}$ and  $\mathbf{r}_i - \mathbf{r}_j$  these  relations are independent of the precise nature of the interactive forces.  In both \eqref{eq:trans} and \eqref{eq:rot}, we also note that it is necessary to evaluate the horizontal flow at the true particle position $\mathbf{r}_i = \wh{\bf r}_i + \bs{\eta}_i $ on the surface.  This subtlety is crucial to find the slow co-rotation of the raft. 

We formulate a continuum limit for a circular raft of radius $a$, composed of many uniformly distributed particles. In the absence of flow, we suppose that the raft is centered on the origin of the cylinder. We then have
\bse \label{eq:transetrot3}
\ba
\frac{d \, \wh{\bf r}_g}{dt}  &=&  \frac{1}{\pi a^2}   \iint_{\mc{D}(t) } \wh{\bf u}(\wh{\bf r} + \bs{\eta} ,t)  \, d^2 \wh{\bf r}  \label{eq:trans3}  \\
 \omega  & =&  \frac{2}{\pi a^4}     \iint_{\mc{D}(t)}  [ (\wh{\mathbf{r}} - \wh{\mathbf{r}}_g )   \times  \wh{\bf u}(\wh{\bf r}+ \bs{\eta} ,t)  ] \cdot \mathbf{e}_z  \, d^2 \wh{\bf r}  \label{eq:rot3}
\ea
\ese
Here we denote $\mc{D}(t)$ is the domain where $ || \wh{\mathbf{r}}  - \wh{\mathbf{r}}_g || \leq a$. It is useful to rewrite the integrals of \eqref{eq:transetrot3} using a translated coordinate system, centered on the raft. There we have 
\bse \label{eq:transetrot3p}
\ba
\frac{d \, \wh{\bf r}_g}{dt}   &=&  \frac{1}{\pi a^2}   \iint_{{\mc{D}} }\wh{\bf u} ( \wh{\bf r}  + \wh{\bf r}_g + \wt{\bs{\eta}} ,t)  \, d^2 \wh{\bf r}  \label{eq:trans3p}  \\
 \omega  & =&  \frac{2}{\pi a^4}     \iint_{{\mc{D}}}  [ \wh{\mathbf{r}}    \times  \wh{\bf u}( \wh{{\bf r}}  + \wh{\bf r}_g + \wt{\bs{\eta}} ,t) ] \cdot \mathbf{e}_z  \,   \, d^2 \wh{\bf r}  \label{eq:rot3p}
\ea
\ese
Here we denote $\wt{\bs{\eta}} (\wh{\mathbf{r}} , t )  = \bs{\eta} (\wh{\mathbf{r}} + \wh{\bf r}_g , t ) $ and  ${\mc{D}}$ is now a stationary circular domain where  $|| \wh{\mathbf{r}} || \leq a$. Since the flow is small and of order $O(\chi)$, we know that $\wh{\bf r} _g , \bs{\eta} = O (\chi)$ too. This allows us  to use Taylor expansions, to derive explicit formula for ${\bf r}_g  $ and $\omega$ in different orders of $\chi$. 

\section{C. First order motion: gyration}

\begin{figure}
\begin{center}
\includegraphics[scale=.4]{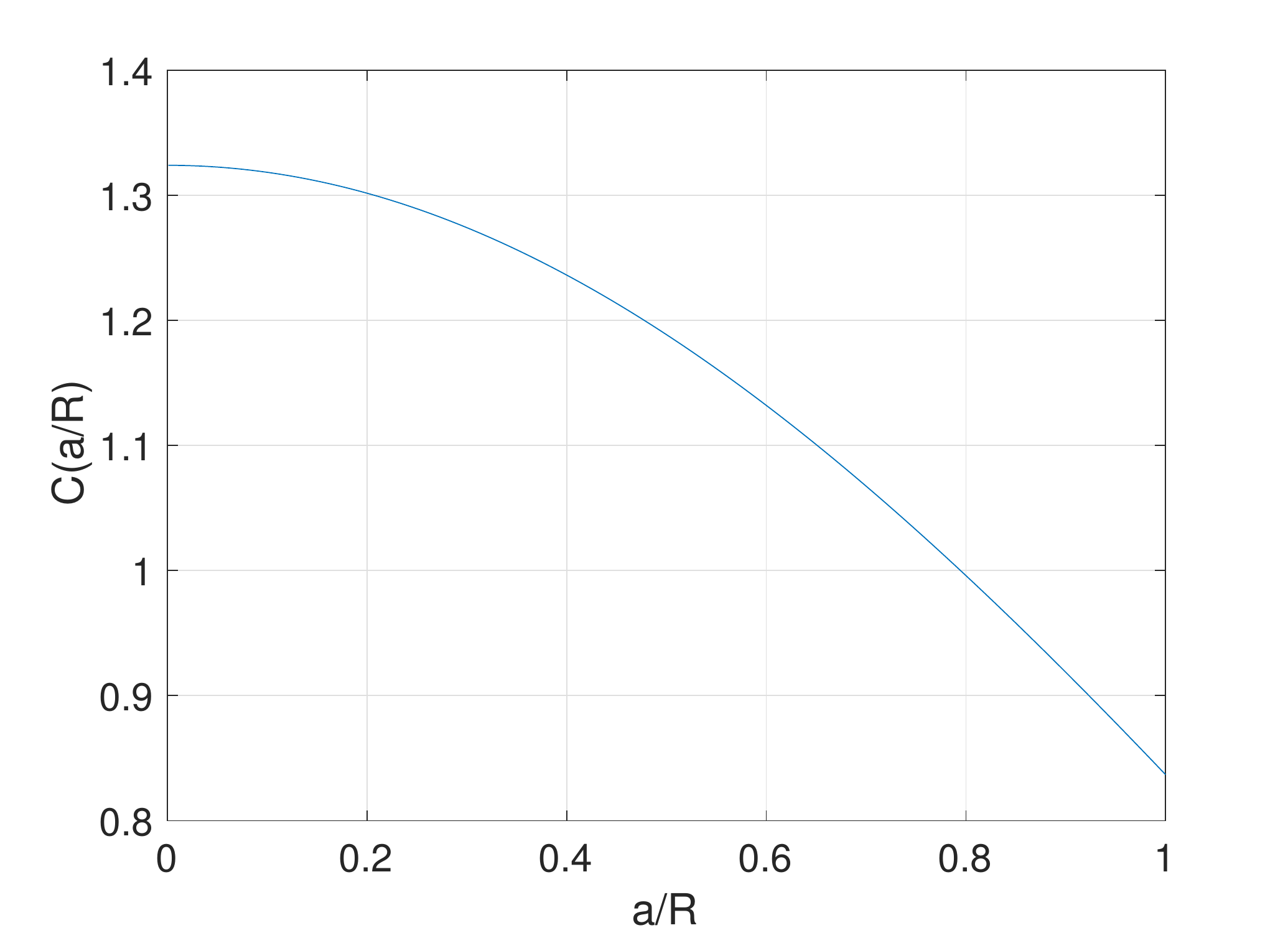}
\includegraphics[scale=.4]{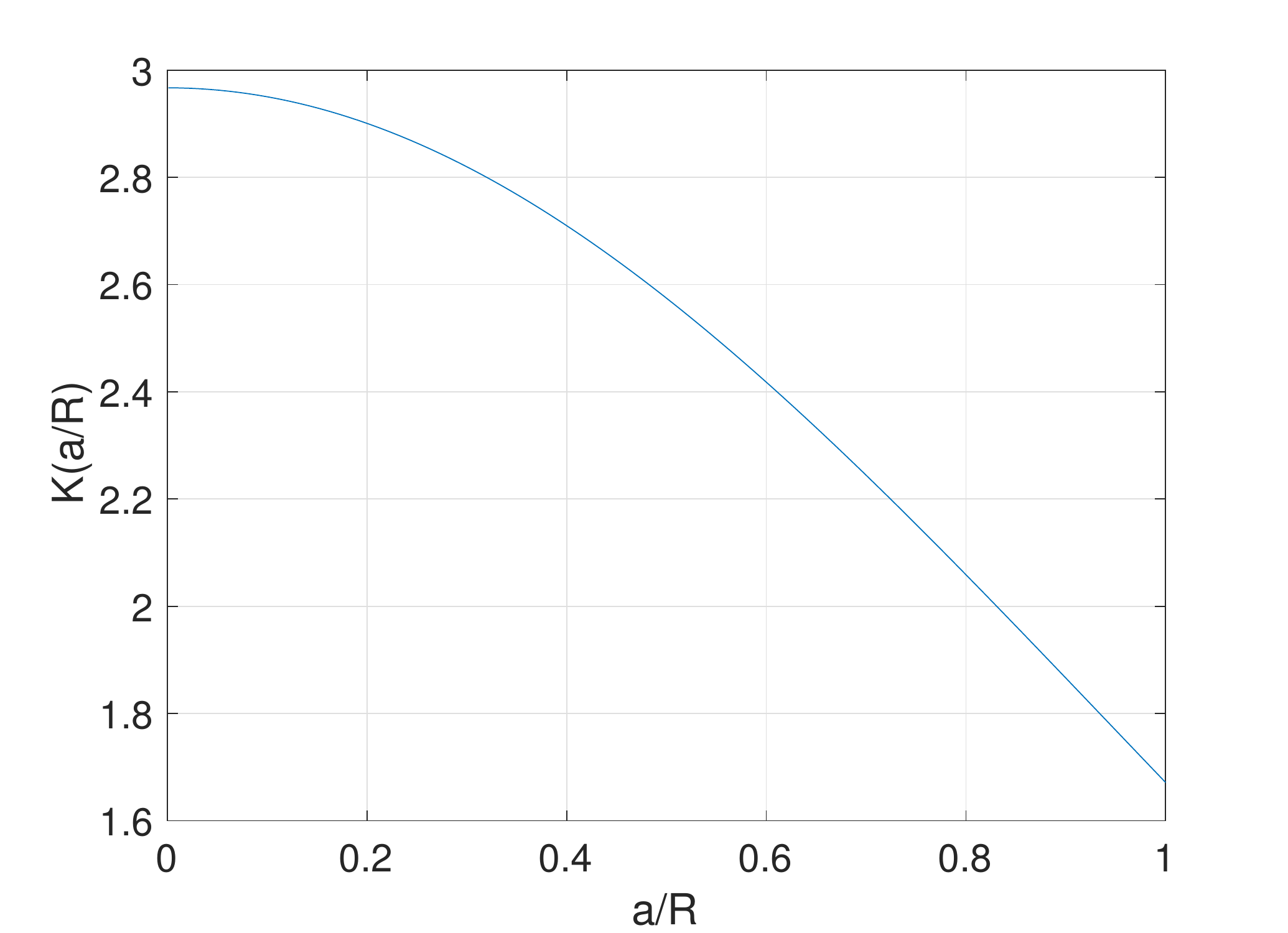}
\caption{ The gyration radius of the raft is $\rho  =  \chi  R \, C (a/R) $ and the rotation speed for the co-rotation is  $\overline{\omega} = \chi^2 \Omega \,  K(a/R)$. Here we show $C(a/R)$ and $K(a/R)$ as functions of the non-dimensional radius  $a/R$ of the raft. We fix $H/R=2.17$ as in the experiment. \label{fig:coef}}
\end{center}
\end{figure}

To obtain a first order approximation for $\wh{\bf r} _g $ and $\omega$, we approximate  
\ba \label{eq:flowO1}
\wh{\bf u} ( \wh{\bf r}  + \wh{\bf r}_g + \wt{\bs{\eta}} ,t)& =& \wh{\bf u} ( \wh{\bf r} ,t )  + O (\chi^2) \nonumber \\
& = &  \wh{\nabla} \phi ( \wh{\bf r}  ,t)   + O (\chi^2)
\ea
in the integrals of \eqref{eq:trans3p} and \eqref{eq:rot3p}. Using the vector identify $\wh{\mathbf{r}} \times \wh{\nabla} \phi  = - \wh{\nabla} \times (\wh{\mathbf{r}} \phi)$ and integration theorems we can simplify the surfaces integrals to contour integrals. 
\bse
\ba
\frac{d \, \wh{\bf r}_g}{dt}   &=&    \frac{1}{\pi a^2 } \int_0^{2 \pi} a \, \phi (a,{\theta},0,t) \,    {\mathbf{e}}_r  \,   d {\theta}    + O (\chi^2) \\
\omega  &=&  \frac{2}{\pi a^4 } \int_0^{2 \pi} a^2 \, \phi (a,{\theta},0,t) \,   \underbrace{( {\mathbf{e}}_r \cdot  {\mathbf{e}}_\phi )}_{=\, 0}   \,   d {\theta}    + O (\chi^2)  =  O (\chi^2).
\ea
\ese
This shows that the raft cannot rotate at first order, we can only have a translational motion. After some calculations, we find  
\be
\wh{\bf r}_g   = \rho \,  \left (  \cos \Omega t  \, \mathbf{e}_x  +   \sin \Omega t   \, \mathbf{e}_y  \right )  + O (\chi^2)  , \label{eq:OGdef} 
\ee
with 
\be
\rho= \chi  R \, \underbrace{  \frac{ 2 R}{a} \frac{1}{(k_1^2 - 1)} \frac{J_1 (k_1 a/R )}{J_1 (k_1)} }_{C (a/R)} . \label{eq:deltadef} 
\ee
We call this motion the gyration of the raft: the center of mass of the raft rotates with time, along with the wave. We denote $\rho$ the gyration radius that scales as $\rho \sim \chi R$. The coefficient of proportionality $C(a/R)$ is shown in figure \ref{fig:coef} as a function of $a/R$. It varies from $1.3$ to $0.8$ for  $a/R$ varying from $0$ to $1$. 

\section{D. Second order motion: co-rotation }

To describe the motion of the raft up to second order, we  approximate 
\ba  \label{eq:flowO2}
\wh{{\bf u}} (\wh{\bf r} + \wh{\bf r}_g + \wt{\bs{\eta}} ,t) &= & \wh{\nabla} \phi \, (\wh{\bf r} ,t)  +  [ ( \wh{\bf r}_g + \bs{\eta} ) \cdot \nabla  ]  \wh{\nabla} \phi   (\wh{\bf r} ,t)   +    \wh{\overline{\bf u}} (\wh{\bf r}) +  \wh{\mathbf{u}}'  (\wh{\bf r} ,t)  + O (\chi^3) 
\ea
in \eqref{eq:trans3p}. In the right hand side, we see a first order Taylor expansion of the potential flow. $\wh{\bs{r}}_g$ can be replaced with \eqref{eq:OGdef} and we can also simplify $\wt{\bs{\eta}} (\wh{\mathbf{r}} , t )  = \bs{\eta} (\wh{\mathbf{r}} , t ) + O (\chi) $. The second order flow correction $ \wh{\overline{\bf u}} + \wh{\mathbf{u}}'$ is also included. We focus on the stationary terms that can induce a slow mean motion. Using bars to denote time-independent fields, we can find that
\ba  \label{eq:flowO2stat}
\overline{\wh{{\bf u}} (\wh{\bf r} + \wh{\bf r}_g + \wt{\bs{\eta}} ,t)} &= &      \wh{\overline{\bf u}} (\wh{\bf r})  + \overline{ ( \bs{\eta}  \cdot \nabla )  \wh{\nabla} \phi  } \, (\wh{\bf r} )   + O (\chi^3)   \\
&=&     \wh{\overline{\bf u}} (\wh{\bf r})  +  \frac{\Omega R^2}{2\, {r}} \left ( \frac{ 2 \chi k_1 }{(k_1^2 -1 )^2 }  \frac{J_1 (k_1 {r} /R) }{J_1 (k_1)} \tanh (k_1 H/R)    \right)^2   \  {\mathbf{e}}_\theta + O (\chi^3) .   \nonumber
\ea
Next to the steady streaming flow for which we have no analytical expression, we find a Stokes drift correction that can be explicitly calculated. 
We admit that $\wh{\mathbf{r}}_g  (t,\tau)$ can have a dependance on a slow time-scale $\tau = (\chi \Omega) ^{-1} $. We then find  that 
\ba
\frac{d\wh{\bf r}_g}{d\tau} &  \simeq&  \frac{1}{\pi a^2}   \iint_{{\mc{D}} }   \left [  \wh{\overline{\bf u}} (\wh{\bf r})   + \overline{ ( \bs{\eta}  \cdot \nabla )  \wh{\nabla} \phi  } \, (\wh{\bf r} )  \right ] \, d^2 \wh{\bf r}  = \bs{0}
\ea
due to axisymetry. The gyration center of an initially centered raft will remain close to the origin on timescales $(\chi \Omega)^{-1}$. For the stationary component $\overline{\omega}$ of the rotation speed \eqref{eq:rot3p} we find up to second order
\be
\overline{\omega} =   \frac{2}{\pi a^4}   \int_0^a \int_0^{2 \pi}   \left [  \wh{\overline{\bf u}} (\wh{\bf r})   + \overline{ ( \bs{\eta}  \cdot \nabla )  \wh{\nabla} \phi  } \, (\wh{\bf r} )  \right ]_\theta    {r}^2 \, d {r} \, d {\theta}  .
\ee
The study of Bouvard et al. [{\rm 4}] suggested that the steady streaming flow has a weak azimuthal component. If we ignore these contributions ($\overline{u}_\theta \approx 0$), we can calculate 
\be
\overline{\omega} \simeq \Omega \chi^2 \underbrace{\frac{R^2}{a^2} \frac{4 k_1^2}{(k_1^2 -1 )^2} \tanh^2(k_1 H/R) \frac{J_1^2(k_1 a/R) - J_0(k_1 a/R)  J_2(k_1 a/R)  }{J_1^2 (k_1)}}_{K(a/R, H/R)} .
\ee
As shown in figure \ref{fig:coef},  $K$ varies from $2.97$ to $1.67$ for  $a/R$ varying from $0$ to $1$ in our set-up with $H/R =  2.17$. A bigger raft rotates slower. The value $2.97$ for very small rafts coincides with Stokes drift rotation speed at the center. In the experiments we found $K \simeq 2$, which is compatible with this result.

\section{E. Boundary layer effects: counter-rotation}

To describe the counterrotating motion of large rafts, we must take into account that such large rafts reach into the boundary layer region while they gyrate. We perform all calculations in the frame attached to the cylinder. We approximate 
\ba  \label{eq:flowO2}
 \wh{\bf u}(\wh{\bf r}+ \bs{\eta} ,t)  =  \wh{\nabla} \phi  - \wh{\nabla} \phi |_{r=R} \,  e^{-(R-r)/\delta}  + O (\chi^2). 
\ea
 in \eqref{eq:rot3}. The effect of potential flow is already known up to second order and induces the co-rotation $\overline{\omega}$. For large rafts, we need to correct the slow rotation speed as
\be
\overline{\omega}_{tot}  = \overline{\omega} + \overline{\omega}_{BL}, 
\ee
where $\overline{\omega}_{BL}$ contains the stationary counter-rotation caused by the boundary layer correction alone. We can calculate  
\be
\omega_{BL}   =  \frac{2}{\pi a^4}    \iint_{\mc{D}(t)}  [ (\wh{\mathbf{r}} - \wh{\mathbf{r}}_g )   \times  ( - \wh{\nabla} \phi |_{r=R} \,  e^{-(R-r)/\delta} )   ] \cdot \mathbf{e}_z  \, d^2 \wh{\bf r}
\ee
and the time-average of this yields $\overline{\omega}_{BL}  $. To evaluate this integral, we parametrize the time-dependent region $\mc{D}(t)$ that is occupied by the raft . If the gyration radius is small compared to the size of the raft ($\rho \ll a$), we can approximate 
\be
{\cal D} (t) \ : \ r \in [0,a + \rho \cos (\theta - \Omega t )] \quad, \quad \theta \in [ 0, 2\pi[  
\ee
up to errors of $O(\rho^2/ a^2 )$. Using the definition  \eqref{eq:OGdef} of the gyration radius $\rho$, we  then express $
\wh{\bf r} - \wh{\bf r} _g$ in cylindrical components to find 
\ba
\omega_{BL} &\simeq & -  \frac{4 \Omega R \chi }{\pi a^4 (k_1^2 -1)} \int_0^{2 \pi}  \!\!\! \int_{0}^{a + \rho \cos (\theta - \Omega t)}  \hspace*{-1.3cm} \left ( r - \rho \cos (\theta - \Omega t) \right )  \cos (\theta - \Omega t) \,  e^{(r-R)/\delta} \, r \, dr \, d \theta  . 
 \ea
 Due to the presence of the exponential factor, the integrandum rapidly decays away from the boundary $r=R$, which allows some simplifications. We introduce a change of variables $s = (r-R)/ \delta$ and approximate the bound $r=0$ by $s = -R / \delta \rightarrow - \infty$. In the integrandum, we also approximate all other occurrences of $r\simeq R$. 
Integration over $s$ is then very simple, giving
\ba
&& \omega_{BL} = -  \frac{4 \Omega R^2 \delta \chi }{\pi a^4 (k_1^2 -1)} \,  e^{(a-R)/\delta}   \int_{0-\Omega t}^{2 \pi- \Omega t}   \left  (R - \rho \cos \wt{\theta}  \right )   \cos \wt{\theta} \, e^{(\rho/\delta)  \cos  \wt{\theta}} \, d \wt{\theta} \nonumber 
\ea
with $\wt{\theta} = \theta - \Omega t$. This integral can be evaluated analytically in terms of modified Bessel functions $I_m$ as we have elementary integrals
\bse
\ba
\int_{0 - \beta}^{2 \pi - \beta} e^{\zeta \cos \wt{\theta} }  \, d \wt{\theta} &=& 2 \pi I_0 (\zeta) \\
\int_{0 - \beta}^{2 \pi - \beta} \cos  \wt{\theta} \,  e^{\zeta \cos \wt{\theta} }  \, d \wt{\theta} &=&  2 \pi I_1 (\zeta)  \\
\int_{0 - \beta}^{2 \pi - \beta} \cos^2  \wt{\theta} \,  e^{\zeta \cos \wt{\theta} }  \, d \wt{\theta} &=&  \pi (I_0 (\zeta)  + I_2 (\zeta) )
\ea
\ese
$\forall \beta \in \Bbb{R}$. The first relation is well known. The second and third relation can be obtained by deriving the first relation with respect to $\zeta$ and using recurrence relations of modified Bessel functions. With this, we obtain 
\ba
 \omega_{BL} &=&  - \chi \Omega  \frac{8  \delta R^3}{ a^4 } \frac{e^{(a-R)/\delta}}{(k_1^2 -1) }   \left \{    I_1 \left (\frac{\rho}{\delta} \right ) - \frac{\rho}{2R} \left [  I_0 \left (\frac{\rho}{\delta} \right ) +    I_2 \left (\frac{\rho}{\delta} \right ) \right ] \right \} . \nonumber
\ea 
We notice that $ \omega_{BL}$ is time-independent and since we have $\rho \ll R$, the first term proportional to $I_1 (\rho/\delta) > 0$ dominates. Therefore, we can expect a counter-rotation. It is useful to remember that this formula only makes sense when $\rho, \delta \ll R$, $\rho + a \leq R $ and when the wave-magnitude remains small.  
In the limit  $\rho \ll \delta$, we can use a Taylor expansion $I_1 (z)  \simeq  z / 2$ and with $\rho \ll R$ we can ignore the contributions from $I_0$ and $I_2$. This then yields
\ba
 \omega_{BL} &\simeq&  - \chi^2 \Omega  \frac{4  R^4}{ a^4 } \frac{e^{(a-R)/\delta}}{(k_1^2 -1) }  C(a/R) , 
   \nonumber
\ea 
with $C(a/R)$  defined in \eqref{eq:deltadef}. We notice that the counter-rotation is of order $\sim \chi^2$ just as the co-rotation.

\end{document}